\documentclass[a4paper, 11pt]{article}
\usepackage[affil-it]{authblk} 
\usepackage{etoolbox}
\usepackage{lmodern}
\usepackage[a4paper]{geometry}
\usepackage[in]{fullpage}
\usepackage[utf8]{inputenc}
\usepackage{amsmath}
\usepackage{amssymb}
\usepackage{cite}
\usepackage{lipsum}
\usepackage{amsfonts}
\usepackage{graphicx}
\usepackage[dvipsnames]{xcolor}
\usepackage[colorlinks]{hyperref}
\hypersetup{linkcolor=blue,citecolor=blue,urlcolor=red}

\usepackage{indentfirst}

\newcommand{\ap}{\alpha'}

\title{String Cosmology backgrounds from Classical String Geometry}

\makeatletter
\patchcmd{\@maketitle}{\LARGE \@title}{\fontsize{16}{19.2}\selectfont\@title}{}{}
\makeatother

\author[1]{Heliudson Bernardo\footnote{\href{mailto:heliudson@hep.physics.mcgill.ca}{heliudson@hep.physics.mcgill.ca}}}
\author[1]{Robert Brandenberger\footnote{\href{mailto:rhb@hep.physics.mcgill.ca}{rhb@hep.physics.mcgill.ca}}}
\author[2]{Guilherme Franzmann\footnote{\href{mailto:guilherme.franzmann@su.se}{guilherme.franzmann@su.se}}}

\affil[1]{Department of Physics, McGill University,\protect\\ Montreal, QC, H3A 2T8, Canada}
\affil[2]{Nordita,
KTH Royal Institute of Technology and Stockholm University, 
Roslagstullsbacken 23, SE-106 91 Stockholm, Sweden}

\date{\vspace{-5ex}}

\begin{document}

\maketitle


\begin{abstract}

We introduce a very early universe model based on the thermodynamics of a gas of closed strings in a background which is non-perturbative in $\ap$. Upon considering the fully $\alpha'$-corrected equations extended to include certain anisotropic cosmological backgrounds, we describe the evolution of the system in three different stages parametrized by the gas' equation of state. Using standard string thermodynamical arguments, we start with an isotropic 10-dimensional universe inside the string scale and evolve it towards a universe with four large spacetime dimensions and six stabilized internal dimensions in the Einstein frame. 

\end{abstract}


\tableofcontents

\section{Introduction}
\label{sec:intro}

\indent The $\Lambda$CDM model is quite successful. Relying solely on six free parameters, it is able to account for most of the current cosmological data \cite{Akrami:2018vks}, which has become abundant for the last 30 years. It  provides a description of the evolution of the Universe that extends from a fraction of a second to its current age, around $13.8$ billion years \cite{Aghanim:2018eyx}.
 
An attachment to the $\Lambda$CDM model is the inflationary paradigm for the very early universe. Inflation \cite{Brout:1977ix,Sato:1980yn,Starobinsky:1980te,Guth:1980zm,Linde:1981mu,Steinhardt:1982kg,Vilenkin:1983xq} postulates a phase of accelerated expansion in the early universe that explains why the Universe we live in seems to be so spatially flat, so large and nearly homogeneous. It also explains how the small fluctuations in the Cosmic Microwave Background (CMB) are generated and why they are almost scale-invariant, and therefore it also explains how structures such as galaxies and galaxy clusters have been formed in our Universe. However, inflation does not explain away all the problems. Both the $\Lambda$CDM model and inflation rely on General Relativity (GR), which is shown to be unavoidably singular in the very early universe considering these models' matter content \cite{Hawking:1969sw,Borde:1993xh,Borde:2001nh,Yoshida:2018ndv}. It is expected that only a fully-fledged theory of Quantum Gravity (QG) could yield a non-singular cosmology, thus explaining what really happens to the spacetime close to diverging curvature regions.

String Theory is one of the most promising candidates for a QG theory. Among its successes, the theory provides a possible framework for unifying all the known interactions of nature. One of the main advantages to consider strings as being fundamental instead of point particles is the fact that the singularity theorems may be avoided. This is easy to understand intuitively, since as the energy scale gets higher, the energy can flow into the additional degrees of freedom present due to the extra dimensionality of the string.

In fact, not only has String Theory new degrees of freedom, it also contain new symmetries. Particularly, on compact manifolds, strings also have winding modes, besides the quantized momentum modes, that correspond to strings wound in closed cycles \cite{Polchinski:1998rq}. Due to the existence of these different types of modes, toroidal compactifications present a new symmetry: T-duality \cite{Giveon:1994fu}. This symmetry implies that physics in geometries with characteristic radius $R$ is equivalent to physics in geometries with characteristic radius $l_s^2/R$, where $l_s$ is the string length. 

Furthermore, it is worth noting that a thermodynamical treatment of a gas of strings also obeys this symmetry, which can be seen from the thermal partition function of a gas of closed strings in a toroidal background \cite{PhysRevD.40.2626}. This implies that the temperature $T(R)$ remains finite as the torus' radius, $R$, runs from $0$ to $l_s$ while considering the total entropy to be constant. Moreover, if the gas of strings contains a large entropy, then, for a wide range of values of $R$ on either side of the string scale, $T(R)$ hovers just below the Hagedorn temperature $T_H$, the maximal temperature for a gas of closed strings \cite{Hagedorn:1965st}.

Given that thermal effects may be important for realistic cosmological backgrounds, the above considerations gave rise to the String Gas Cosmology (SGC)\footnote{Recently the dynamics of SGC has been embedded into a more general proposal called \textit{Emergent scenario} \cite{Brandenberger:2019jbs}.} scenario \cite{Brandenberger:1988aj}, (see also \cite{Kripfganz:1987rh}) according to which the spacetime geometry is locally $\mathbb{R}\times\mathbb{T}^9$ and the universe emerges from a phase in which matter is made of a gas of strings with temperature close to the Hagedorn temperature, while the T-duality symmetry in the matter sector is unbroken. It was postulated that this phase is quasi-static in the sense that the scale factor in the Einstein frame (EF) is nearly constant. Later, it was shown that thermal fluctuations of the string gas lead to a nearly scale-invariant spectrum of cosmological perturbations with a small red tilt \cite{Nayeri:2005ck} for the scalar modes  and  a slight blue tilt \cite{Brandenberger:2006xi, Brandenberger:2014faa} for tensor modes. The fluctuations are Gaussian and have Poisson-suppressed non-Gaussianities on large scales \cite{Chen:2007js}. Hence, SGC yields an alternative to the cosmological inflationary paradigm for explaining the origin of structure in the universe (see e.g. \cite{Brandenberger:2011et,Brandenberger:2008nx,Battefeld:2005av} for reviews of SGC).

As studied in \cite{Patil:2004zp, Patil:2005fi, Watson:2003gf, Watson:2004aq}, size moduli of the extra spatial dimensions are naturally stabilized at the string scale by the interplay between momentum and winding modes. Similarly, shape moduli of the extra dimensions can be stabilized by stringy effects \cite{Brandenberger:2005bd}. Non-perturbative effects like gaugino condensation can be used to stabilize the dilaton \cite{Danos:2008pv} without interfering with the stabilization of the other moduli. This non-perturbative mechanism then leads to supersymmetry breaking at the string scale \cite{Mishra:2011fc}. The key open issue in SGC is to justify the assumption that the EF scale factor is in fact nearly constant in the high temperature phase. If we were to use Einstein gravity, we would not obtain an almost constant scale factor in a phase of high string gas energy density. 

However, the Einstein equations are clearly not the correct equations to use for the background dynamics since they are inconsistent with the T-duality symmetry of String Theory. Pre-Big-Bang Cosmology \cite{Gasperini:1992em} (see also \cite{Tseytlin:1991xk}, and \cite{Gasperini:2002bn} for a review) is an attempt to study early universe cosmology in the context of dilaton-gravity where there is a scale factor duality symmetry between solutions. However, the static phase required by SGC is not a solution of the equations, and even if it was it would not be justified since such equations are not valid anymore for high energy densities. 

The dilaton-gravity equations are actually low energy equations for the bosonic sector of the supergravity theory for the background (massless) fields of superstring theories, once we turn off the all the fluxes. In fact, the massless Neveu-Schwarz (NS-NS) sector is universal for all $10$-dimensional superstring theories and has the same action for closed superstrings \cite{Polchinski:1998rr}. In applications to cosmology, such equations are typically sourced by the energy-momentum tensor of a perfect fluid \cite{Gasperini:2007zz}. For a gas of strings, the energy-momentum tensor has exactly this form with an equation of state (EoS) that depends on the modes that dominates the gas: for compact directions with size smaller (greater) than the string length, winding (momentum) modes are energetically favorable \cite{Brandenberger:1988aj}.  

If we are after solutions with high energy density, such as during the static phase in the EF of SGC, we need to correct the bosonic NS-NS sector of the supergravity action with higher order operators. These operators are associated with $\alpha'$ and $g_s$ corrections. The former are related to the string length, given by $l_s \equiv \sqrt{\alpha'}$ which sets the string scale, thus present even at classical level, while the latter are due to string interactions and account for quantum corrections. They correspond to the $2$-dimensional sigma model and spacetime perturbative expansions, respectively. Having the set of fully corrected equations is one of the most desirable achievements in String Theory, as it could be used to answer all sorts of non-perturbative and phenomenological questions.

An interesting point of view is that due to the extensive nature of its fundamental constituents, String Theory gives rise to a new kind of geometry at the non-perturbative level, a \emph{string quantum geometry} \cite{Greene:1996cy}. In the limit $g_s \to 0$ and $\alpha'/\mathcal{R}^2 \to 0$ (where $\mathcal{R}$ is the characteristic radius of spacetime curvature) we are back to Einstein theory plus classical fields, while at any given order in both expansions there are corrections to this limit. Note that these limits are not completely independent, since the string coupling is not a free parameter, being fixed by the dilaton's vacuum expectation value, $g_s = e^{\langle \phi \rangle}$. However, if the equations admit solutions with a small string coupling, then we can neglect the quantum corrections to leading order while keeping all $\alpha'$-corrections in the non-perturbative regime, which gives rise to the \emph{classical string geometry} limit, i.e., the geometry of the tree-level String Theory.

Although it is expected that the final equations are background invariant \cite{Hohm:2018zer}, significant progress has been made recently for purely time-dependent backgrounds at tree-level. This was due to the fact that for such backgrounds  there is a non-compact symmetry acting in the field space \cite{Maharana:1992my}. Indeed, for a cosmological ansatz in $D = d+1$ dimensions, the scale factor duality \cite{Meissner:1991zj} is a particular discrete transformation within a global O$(d,d)$ group \cite{Gasperini:1991ak}. Restricted to the lowest order terms, a duality covariant formalism was established in \cite{Gasperini:1991ak}, including the energy momentum tensor of a gas of strings, that was shown to transform covariantly under the O$(d,d)$ group. Moreover, in \cite{Sen:1991zi} it was shown that the O$(d,d)$ symmetry should be present to all orders in $\alpha'$ and, in fact, it was shown in \cite{Meissner:1996sa} that although the first corrections modify the duality transformation, there are field variables in which they remain unchanged. Assuming that to be the case at any order, all possible corrections were classified in \cite{Hohm:2019jgu} for the vacuum case. The formalism was extended to include matter couplings through an O$(d,d)$ invariant matter action in \cite{Bernardo:2019bkz}, establishing the $\ap$-Cosmology framework within which perturbative and non-perturbative solutions were found. In \cite{Bernardo:2020zlc} such solutions were shown to hold even with a non-trivial dilatonic charge, and their stability under homogeneous perturbations was studied. 

In the following sections, we propose an early universe cosmological scenario based on these solutions. It starts off with ten dimensions where nine spatial dimensions are smaller than the string length and evolves such that at the end we have four large spacetime dimensions while the other six spatial dimensions remain stabilized around the string length. This is realized after considering a gas of strings sourcing the equations assuming the expected evolution of the equation of state for a gas of strings in the most natural way \cite{Brandenberger:2018xwl} and then solving the dynamics in three stages. Surprisingly, the $\alpha'$-corrected equations support a static phase in the Einstein frame as postulated by SGC, though in ten dimensions.  

The outline of the paper is as follows. In Section \ref{summary}, we summarize the construction and heuristics of the model. In Section \ref{technical}, we discuss technical details, in particular how we can get $4$-dimensional equations from $\alpha'$-Cosmology after having extended the framework to include a certain class of anisotropic cosmological backgrounds. The quantitative aspects of the model are introduced in Section \ref{the_model}, where both the dynamics in the String frame and in the Einstein frame are discussed. Then we conclude in Section \ref{Conclusion}.

\section{Summary of the Model}\label{summary}

In SGC, the thermodynamics of a gas of strings in a $(d+1)$-dimensional compact space can be separated in 3 types of equations of state (EoS) assuming a barotropic perfect fluid, such that $p = w \rho$: a winding EoS, with $w = -1/d$; a radiation EoS, $w = 1/d$; and a pressureless one, with $w = 0$. Indeed, for a non-interacting isotropic gas of strings on an isotropic toroidal background $\mathbb{T}^{d}$ winding modes are energetically favorable if the radius of the torus is smaller than the string length, so that the fluid is dominated by these modes and has a winding EoS. On the other hand, momentum modes dominate when the radius is greater than the string length such that the fluid has a radiation EoS in this case. Close to the T-duality self-dual radius $\sqrt{\alpha'}$, both modes contribute with the same magnitude to the pressure, but with opposite signs, giving rise effectively to a dust-like EoS, since the oscillatory modes which are also excited around the self-dual radius yield a pressureless fluid as well \cite{Battefeld:2005av}.

Note that in order to calculate how each string state contributes to the energy and pressure of the string gas, the mass spectrum of a single string in a static toroidal spacetime is used. In SGC, an adiabatic approximation is assumed, such that we can approximate the spectrum in a cosmological spacetime by simply promoting the radius of the torus to be the time dependent scale factor \cite{Battefeld:2005av}. In the following, we use the results obtained from the adiabatic approximation, in particular the equations of state described above, even though the background is an expanding Friedmann-Lema\^itre-Robertson-Walker (FLRW) cosmology with Hubble parameter close to the string scale. The justification comes from T-duality, since once it holds to all orders in $\alpha'$, there should be winding and momentum modes among the states. Thus, even for the full spectrum in the time-dependent background, we expect these modes to dominate the string gas states. 

In the String frame (SF), we assume an initial high density string gas phase on a cosmological spacetime that has the topology of $\mathbb{R}\times \mathbb{T}^9$ with all spatial directions compatified on a $9$-dimensional torus with radius smaller than the string length. As discussed in previous paragraphs, the string gas starts off with a winding EoS. Now, given that the energy density is closer to the string scale, the $\alpha'$-corrected cosmological equations of \cite{Bernardo:2019bkz, Bernardo:2020zlc} should be the ones to rule the background evolution, which is expected to be non-perturbative in $\alpha'$. It was shown that there are non-perturbative $d$-dimensional de Sitter (dS) solutions, $H(t)=H_0$ (in String frame) with constant equation of state, $w=w_0$, and the dilaton's evolution completely parametrized by these two constants (see \eqref{eq:dot_dilaon_for_dS}). Thus, the natural solution for this initial stage is a compactified $dS_{10}$ solution with a winding equation of state, with all directions expanding until their physical radius become of the order of $\sqrt{\alpha'}$. This stage corresponds to a static phase in the Einstein frame as described in Section \ref{EF_stage_1}.

As the background approaches a characteristic length equal to the string size, the string gas fluid ceases to have a winding EoS, since oscillatory and momentum modes start to get excited. Thus, the EoS evolves towards zero as the physical radii get closer to the string scale. This establishes the second stage of the dynamics. There is an important caveat here: as the EoS approaches zero, the geometry departures from being isotropic in all spatial directions and it divides into two independent isotropic sectors: an \emph{internal} six-dimensional one, for which the EoS associated with these dimensions stops evolving as it reaches zero, and an \emph{external} three-dimensional one, for which the EoS keeps evolving towards a radiation EoS. This happens because the winding modes can annihilate completely only in the latter sector, while in the former they remain existing and helping to stabilize the internal EoS together with the momentum modes \cite{Brandenberger:1988aj}. In Section \ref{sec:non_isotropic_solutions}, we show explicitly that there are solutions with static directions with $p_i=0$ and $\dot{w}_i = 0$ so that the balance between winding and momentum modes at the string scale can potentially stabilize all the internal directions, as previously suggested by \cite{Watson:2003gf,Patil:2004zp}. This is the end of the dynamics of the \emph{internal} directions in the SF, while in the EF they start to contract as described in Section \ref{EF_stage_2}. 
 
Since the winding modes completely decay into momemtum modes in the external sector, the EoS continues evolving until becoming a radiation EoS, which allows the external directions to remain dynamical. This, together with the freezing of the internal sector, is the SGC mechanism for generating a three-dimensional cosmology from a 10-dimensional one. After the radiation EoS is settled, we enter the third stage of the model, where we have an anisotropic cosmology with six static spatial \emph{internal} directions with $w =0$ and three evolving \emph{external} directions with $w= 1/3$. The relevant non-perturbative solution is now locally $\text{dS}_4 \times \mathbb{T}^6$ (in the String frame) with a rolling dilaton whose velocity is determined by the evolution of the external directions (as explictly shown in Section \ref{SF_stage_3}).

Meanwhile, the dilaton has been evolving so far approximately linearly with time. Although it is possible to choose the dilaton's initial value such that we reach the third stage in the small string coupling regime, when the internal directions are stabilized and the external ones continues to expand, there is no bound on the dilaton's time evolution. Thus, we eventually enter in the quantum non-perturbative regime, where $g_s = e^{\phi} \sim 1$. From this moment on, as we would like to have a String Theory based model, we cannot fully trust the O$(d,d)$ covariant equations anymore because they do not include $g_s$ corrections. Instead, it is known that non-perturbative effects like D-branes may dominate the theory's spectrum, giving rise for instance to gaugino condensation \cite{Ferrara:1982qs,Affleck:1983rr,Dine:1985rz}. Physically the dilaton should acquire a potential that stabilizes it \cite{Damour:1994ya,Damour:1994zq,Gasperini:2007ar} (see \cite{Danos:2008pv} for dilaton stabilization with gaugino condensation in SGC). Thus, in order to potentially make contact with Standard Big Bang Cosmology, we seek for perturbative solutions with a constant dilaton. In \cite{Bernardo:2019bkz} it was shown that this condition completely fixes the solution to be a perturbatively corrected radiation solution of the graviton-dilaton equations that is known to be determined once a constant dilaton is assumed \cite{Tseytlin:1991xk}. As time goes by, the perturbative corrections get smaller and the solution approaches the lowest order one with a radiation EoS.

After the stabilization of the dilaton, there is no difference between the String and Einstein frames. But during the three stages described above, the dilaton time dependence is fixed by the solutions. That is the reason why it is possible to describe how each stage evolves in the EF. It is important to notice that both frames are equivalent in the sense that any physical observable can be calculated and has the same value regardless of frames \cite{Alvarez:2001qj}. Besides that, describing the EF evolution is useful when trying to make contact with observations. During the first stage, with a winding EoS, the Einstein frame scale factor is constant and we have a static $10$-dimensional phase as a solution of the non-perturbative equations, in contrast with the $4$-dimensional static phase postulated by SGC. In the second and third stages, the dilaton's evolution is independent of the static internal directions that in  the EF corresponds to contracting dimensions, while the external directions first undergo accelerated expansion and then later expand as a radiation dominated universe.

\begin{figure}[h!]
    \centering
    \includegraphics[scale=.73]{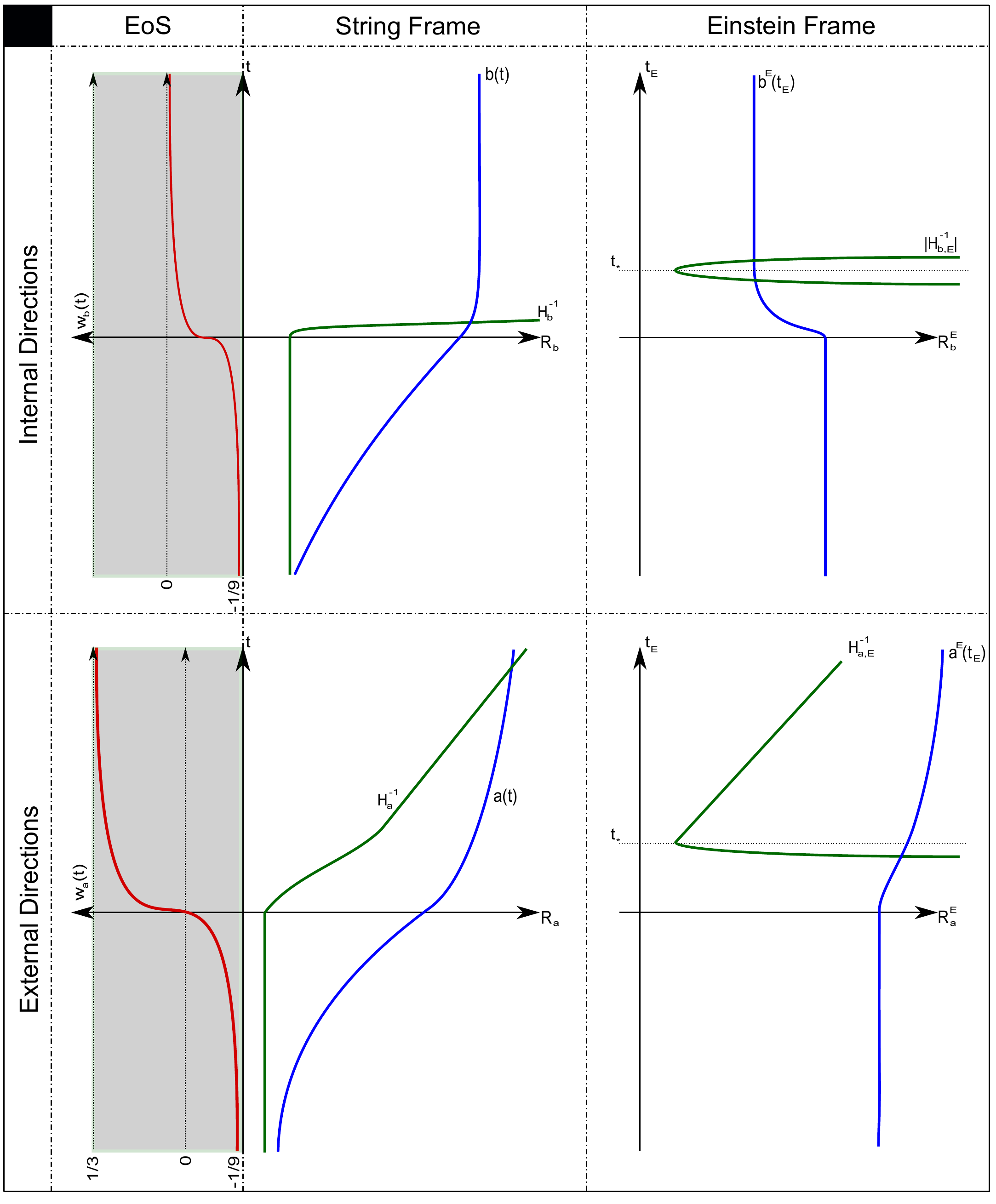}
    \caption{The dynamics of the model is shown as a whole, which is later separated into three different stages in Section \ref{the_model}. On the left, the evolution of the equation of state is plotted as a function of time for both the internal and external directions. In the center, the time evolution of the scale factor and the Hubble radius can be seen in the String frame while on the right they are shown in the Einstein frame. The dilaton stabilizes at $t_*$. The evolution is non-singular in both frames.}
    \label{fig:model}
\end{figure}

The model is summarized in Figure \ref{fig:model}, where the Hubble radius of the internal and external directions and the EoS are schematically plotted as a function of time. Quantitative details about the stages can be found in Section \ref{the_model}. Finally, it is important to emphasize that prior to $\ap$-Cosmology, there were no equations that could describe the dynamics of the model as elucidated above. In particular, the existence of static solutions in the EF for a winding EoS and the static solutions for the internal directions in the SF with a pressureless EoS, which were essential for the model, are  here derived for the first time using the framework discussed here.

\newpage

\section{$\alpha'$-Cosmology: time-dependent backgrounds from classical string geometry}\label{technical}

\subsection{Review of $\alpha'$-Cosmology}

In \cite{Hohm:2019jgu,Bernardo:2019bkz}, it was shown that the action for a purely time-dependent $D = d+1$-dimensional string background, including a matter sector, with metric $G_{00} = -n^2(t)$, $G_{0i} = 0$, $G_{ij} = g_{ij}(t)$, Kalb-Rammond field $B_{00}=0 = B_{0i}$, $B_{ij} = b_{ij}(t)$ and dilaton field $\phi(x) = \phi(t)$ can be written in a O$(d,d)$ invariant form as 
\begin{align}
    S &= \frac{1}{2\kappa^2}\int d^d x dt n e^{-\Phi}\left[-(\mathcal{D}\Phi)^2 + X(\mathcal{DS})\right] + S_{m}[\Phi, n, \mathcal{S}, \chi],
\end{align}
where $\Phi \equiv 2\phi - \ln \sqrt{\det g}$ is a O$(d,d)$ scalar called the shifted dilaton, $\chi$ represents the matter sector and the O$(d,d)$ scalar function $X(\mathcal{DS})$ depends only on the first time derivative ($\mathcal{D} \equiv 1/n \partial_t$) of the $2d\times2d$ matrix
\begin{equation}
    \mathcal{S} \equiv \eta \mathcal{H} =
    \begin{pmatrix}
    0 & 1 \\
    1 & 0 
    \end{pmatrix}
    \begin{pmatrix}
    g^{-1} & -g^{-1}b \\ 
    bg^{-1} & g - bg^{-1}b
    \end{pmatrix} = 
    \begin{pmatrix}
    bg^{-1} & g- bg^{-1}b \\
    g^{-1} & - g^{-1}b 
    \end{pmatrix},
    \end{equation}
where we use a basis in which the O$(d,d)$ metric $\eta$ has an off-diagonal form and $\mathcal{H} \in \text{O}(d,d)$ acts like a generalized metric. As $\mathcal{H}$ is an element of the duality group, $\mathcal{S}$ is a constrained field satisfying $\mathcal{S}^2 =1$. Since the gravitational coupling $\kappa^2$ is factorized, terms in the integrand should have mass dimension $2$. Given that the function $X$ is invariant under duality transformations, it can be written as a sum of traces of the matrix $\mathcal{S}$. Thus, at a given order $k -1$ in $\alpha'$, there could be two types of dimension $2$ operators in the action: single-trace and multi-trace ones, with respective forms,
\begin{equation}
    \alpha'^{k-1}\text{tr}\left((\mathcal{DS})^{2k}\right), \quad \alpha'^{k-1}\prod_{i=1}^{j}\text{tr}\left((\mathcal{DS})^{2l_i}\right),  
\end{equation}
where the set $\{l_i\}$ is constrained in order for the operator to have dimension 2, $l_1+\dots+ l_j = k$. Moreover, as shown in  \cite{Hohm:2019jgu}, by redefining $n^2(t)$ we can set to zero any multi-trace operator containing factors of $\text{tr}(\mathcal{DS})^2$, so $l_i \neq 1$. Thus, the number of multi-trace operators at the $(k-1)$th order is the number of partitions of $k$ that does not include 1, i.e. $j=1,\dots, p(k)-p(k-1)$, where $p(k)$ is the number of partitions of $k$. Note that the lowest order action is obtained by truncating the corrections to $k=1$, and this gives a single-trace operator proportional to $\text{tr}((\mathcal{DS})^2)$.

It was noticed in \cite{Hohm:2019jgu} that with a flat FLRW ansatz for the metric and vanishing 2-form field, the multi-trace operators contribute in the same manner as the single-trace ones and thus they only renormalize the numerical coefficients of the latter. Hence, for this specific ansatz, we can neglect the multi-trace operators such that the action has the form
\begin{equation}\label{singletraceaction}
    S= \frac{1}{2\kappa^2}\int d^d x dt n e^{-\Phi}\left[-(\mathcal{D}\Phi)^2 + \sum_{k=1}^{\infty}\alpha'^{k-1}c_k \text{tr}(\mathcal{DS})^{2k}\right] + S_{m}[\Phi, n, \mathcal{S}, \chi],
\end{equation}
with $c_1 = -1/8$ and the other $c_k$ are generally unknown (they depend on which type of string theory is considered). If $D$ is not equal to the critical dimension $D_c$, there is a term proportional to $e^{-\Phi}(D-D_c)$ in the integrand of the matter action. In the present work, we assume $D = D_c=10$ unless stated otherwise. 

The equations of motion for $\Phi$, $n(t)$ and $\mathcal{S}$ coming from the action (\ref{singletraceaction}) are, respectively,
\begin{subequations}\label{eq:EOM_matter}
    \begin{align}\label{eq:EOM_Phi_matter}
         2\mathcal{D}^2{\Phi} - (\mathcal{D}\Phi)^2 - \sum_{k=1}^{\infty}\alpha'^{k-1}c_k \text{tr}(\mathcal{DS})^{2k} & = \kappa^2 e^{\Phi} \bar{\sigma},\\ \label{eq:EOM_n_matter}
        (\mathcal{D}\Phi)^2 - \sum_{k=1}^{\infty}\alpha'^{k-1}(2k-1)c_k \text{tr}(\mathcal{DS})^{2k} &= 2\kappa^2 \Bar{\rho}e^{\Phi},
  \\ \label{eq:EOM_S_matter}
         \mathcal{D}\left(e^{-\Phi}\sum_{k=1}^{\infty}\alpha'^{k-1}4kc_k\mathcal{S}(\mathcal{DS})^{2k-1}\right) &= -\kappa^2 \eta \Bar{\mathcal{T}}.
    \end{align}
\end{subequations}
The right hand sides (RHS) of these equations are proportional to variations of the matter action. We defined an O$(d,d)$ invariant dilatonic charge $\sigma$ by
\begin{equation}
    \sigma \equiv -\frac{2}{\sqrt{-G}}\frac{\delta S_m}{\delta \Phi},
\end{equation}
while the energy density is given by $n^2(t) \rho = T_{00}$, with
\begin{equation}
    T_{\mu\nu} = -\frac{2}{\sqrt{-G}}\frac{\delta S_m}{\delta g^{\mu\nu}},
\end{equation}
being the energy-momentum tensor of the matter sector. Its spatial components enter in the O$(d,d)$ tensor $\mathcal{T}$ defined by 
\begin{equation}
    \mathcal{\Bar{T}} \equiv \frac{1}{n}\left(\eta\frac{\delta S_m}{\delta \mathcal{S}} \mathcal{S}- \eta \mathcal{S}\frac{\delta S_m}{\delta \mathcal{S}}\right),
\end{equation}
where the bars in the matter variables denotes multiplication by $\sqrt{g}$.

In \cite{Bernardo:2019bkz,Bernardo:2020zlc}, solutions for equations (\ref{eq:EOM_matter}) for the FLRW ansatz
\begin{equation}\label{FLRWansatz}
    ds^2 = - dt^2 + a^2(t)\delta_{ij} dx^i dx^j, \quad b_{ij} = 0, \quad T_{\mu}^{\;\nu} = \text{diag}(-\rho, p, \dots, p),
\end{equation}
were found. Some of these solutions are relevant for the current work. First, there is a non-perturbative class of dS solutions with 
\begin{equation}\label{dSsolution}
    H(t) = H_0, \quad \dot{\Phi} = -\beta H_0,
\end{equation}
where $\beta$ is fixed by the EoS, $p = w\rho$, and the dilatonic charge, assumed to satisfy $\sigma = \lambda \rho$,
\begin{equation}\label{relationbetweenEOS}
    \beta = -\frac{d w}{1 + \lambda/2}.
\end{equation}
Thus, the dilaton evolves as
\begin{equation}
    \dot{\phi} = \frac{dH_0}{2+\lambda} \left(1+w+\frac{\lambda}{2}\right) \label{eq:dot_dilaon_for_dS}.
\end{equation}

There are conditions for the existence of this class of dS solution: the function 
\begin{equation}\label{Ffunction}
    F(H) = 2d \sum_{k=1}^{\infty}(-\alpha')^{k-1}c_k 2^{2k} H^{2k}
\end{equation}
and $H_0$ should be such that (for $\lambda \neq -1,\, 2$)
\begin{equation}
    F'(H_0) = \frac{2d^2w^2H_0}{1+\frac{\lambda}{2}}, \quad 
    F(H_0) = \frac{1+\lambda}{2+\lambda} H_0 F'_{0}.
\end{equation}
While there are sets of $\{c_k\}$ that are not inconsistent with such conditions, the ones coming from String Theory might be incompatible with them. In other words, there are dS solutions (in the SF) in the space of duality invariant theories, but it is not guaranteed that they exist in String Theory\footnote{In \cite{Kutasov:2015eba} a symmetry based argument against the existence of a possible 2d CFT with a ``macroscopic" dS target space was developed. However, we expect the dS radius of our solutions to be of the order of the string length, even though it is not possible to compute the exact relation between $H_0$ and $\sqrt{\alpha'}$ at the moment due to the lack of knowledge on the function $F(H)$ (see \cite{Nunez:2020hxx} for a discussion about how non-perturbative information is necessary to fix $F(H)$). We thank Savdeep Sethi for discussions about this point.}. Nonetheless, no obstruction for such solutions was found after including all $\alpha'$-corrections. For discussions about solutions in the vacuum case, see \cite{Hohm:2019jgu, Wang:2019mwi,Krishnan:2019mkv, Wang:2019kez, Wang:2019dcj}. 

The other relevant solution is a perturbative one. In \cite{Bernardo:2019bkz}, it was shown that the only perturbative solution for a constant dilaton has the form
\begin{align}
    H(t) &= \frac{H_0}{t} + \alpha' \frac{H_1}{t^3} + \alpha'^2 \frac{H_2}{t^5} + \dots,\\
    w(t) &= \frac{1}{d}-32 d  c_2 w_2 \left(\sqrt{\alpha'} H\right)^2 +128 d c_3 w_3 \left(\sqrt{\alpha'} H\right)^4 -512 d c_4 w_4 \left(\sqrt{\alpha'} H\right)^6 + \dots,
\end{align}
where the coefficients $H_1,\,H_2,\dots$ and $w_2,\,w_3,\dots$ depend on the spacetime dimension and the $\{c_k\}$. Thus, it is a solution for any duality invariant theory, in particular for any type of Superstring Theory.
Note that this solution for large times approaches the usual radiation phase of Standard Cosmology.

The solutions above describe $D=d+1$ dimensional isotropic cosmologies, with a single scale factor $a(t)$. In order to discuss more realistic scenarios with a $n$-dimensional compact submanifold, it is useful to have anisotropic solutions, in which the $n$ compact directions evolve differently than the other $d-n$ spatial directions. These solutions cannot be straightforwardly obtained from action (\ref{singletraceaction}) as for them the multi-trace operators cannot simply be taken into account by only redefining the coefficients of the single-trace operators. In the following we show a way around this issue for a particular class of anisotropic cosmological metrics.

\subsection{Anistropic metric in $\alpha'$-Cosmology}

Let us consider a Bianchi type I ansatz for the metric
\begin{align}\label{nonisoansatz}
    ds^2 = -n^2(t)dt^2 + a_i^2(t)\delta_{ij}dx^idx^j,
\end{align}
and vanishing 2-form field, $b_{ij} = 0$. For this particular case, we have 
\begin{equation}\label{Sdot}
    \mathcal{D S} = 2 \mathcal{K} \mathcal{J},
\end{equation}
where $\mathcal{K}$ is a $2d\times 2d$ diagonal matrix constructed with $d$ Hubble parameters, $H_i \equiv  \mathcal{D} \ln a_i$,
\begin{equation}
    \mathcal{K} = \begin{pmatrix}
    H_i & 0 \\
    0 & H_i 
    \end{pmatrix},
\end{equation}    
and $\mathcal{J}$ is defined by 
\begin{equation}
    \mathcal{J} =\begin{pmatrix}
    0 & g \\
    -g^{-1} & 0 
    \end{pmatrix},
\end{equation}
and it squares to minus the identity, $\mathcal{J}^2 = -\mathcal{I}$.

Using (\ref{Sdot}), we evaluate typical single and multi-trace operators for the anisotropic ansatz to be
\begin{equation}\label{singleandmultitraceop}
    \text{tr}\left((\mathcal{DS})^{2k}\right) = (-1)^k 2^{2k+1}\sum_{i=1}^d H_i^{2k}, \quad \prod_{i=1}^{j}\text{tr}\left((\mathcal{DS})^{2l_i}\right) = (-1)^k 2^{2k+1}\prod_{i=1}^{j}\sum_{q=1}^{d}H_q^{2l_i},
\end{equation}
where the constraint $l_1+\dots + l_j=k$ was used. We see that in the isotropic case, $H_i = H \; \forall \; i$, they have the same structure and so contribute in the same form to the equations of motion. 

Now, let us suppose we have $(d-n)$-directions with the same scale factor, i.e. $a_i(t) = a(t)$ for $i = 1,\dots, d-n$, and $n$ static directions with $H_i = 0$ for $i=d-n+1,\dots,d $. In this case, we have 
\begin{equation}\label{singleandmultitraceopunderansatz}
    \text{tr}\left((\mathcal{DS})^{2k}\right) = (-1)^k 2^{2k+1}(d-n)H^{2k}, \quad \prod_{i=1}^{j}\text{tr}\left((\mathcal{DS})^{2l_i}\right) = (-1)^k 2^{2k+1}(d-n)^j H^{2k},
\end{equation}
and so, for this particular case, the $(k-1)$th order multi-trace operators contribute in the same way as the single-trace ones, they merely shift the coefficient $c_k$ of the latter. Hence, \emph{for $n$ static directions and $(d-n)$ isotropic directions}, we can neglect the multi-trace operators and use action (\ref{singletraceaction}) to get the equations of motion for the $(d-n)$ dynamical scale factors. Therefore, the single-trace action can also be useful for finding some anisotropic solutions.

Note that in this calculation we have assumed that $n$ directions were static. Thus, if we plan to use it later we will have to invoke good reasons why that should be the case. Nevertheless, we can use the full Bianchi type I ansatz (\ref{nonisoansatz}) in equations (\ref{eq:EOM_matter}) to seek for conditions on the matter sector in order to have $n$ static directions. For that, we evaluate these equations in terms of $H_i$ and the matter variables in the following.

The form of the single-trace operators was already calculated in (\ref{singleandmultitraceop}), and using this result, equations (\ref{eq:EOM_Phi_matter}) and (\ref{eq:EOM_n_matter}) can be written as
\begin{align}
    2 \mathcal{D}^2\Phi - (\mathcal{D}\Phi)^2 + \frac{1}{d}\sum_{i=1}^d F(H_i) &= \kappa^2 e^{\Phi}\bar{\sigma},\\
    (\mathcal{D}\Phi)^2 + \frac{1}{d}\sum_{i=1}^d H_i F'(H_i) - \frac{1}{d}\sum_{i=1}^d F(H_i) &= 2\kappa^2 e^{\Phi}\bar{\rho},
\end{align}
where the function $F$ is as defined in (\ref{Ffunction}). To write equation (\ref{eq:EOM_S_matter}) in components, we assume $T_{i}^{\;j} = p_i \delta_{i}^{\;j}$ and get
\begin{equation}\label{oddT}
    \bar{\mathcal{T}} = \sqrt{g}\begin{pmatrix}
    0 & p_i \delta^i_{\;j}\\
      -p_i \delta_i^{\;j} & 0 \end{pmatrix},
\end{equation}
and thus, the combination $\eta \bar{\mathcal{T}}$ that appears in the right hand side of (\ref{eq:EOM_S_matter}) is 
\begin{equation}
    \eta \bar{\mathcal{T}} = \sqrt{g}\begin{pmatrix}
    -p_i & 0\\
     0 & p_i \end{pmatrix}.
\end{equation}
Moreover, starting from (\ref{Sdot}) one can easily get
\begin{equation}
    (\mathcal{DS})^{2k-1} = (-1)^{k-1}2^{2k-1}\mathcal{H}^{2k-1}\mathcal{J} \implies \mathcal{S}(\mathcal{DS})^{2k-1} = (-1)^{k-1}2^{2k-1}\begin{pmatrix}
    -H_i^{2k-1} & 0\\
     0 & H_i^{2k-1} \end{pmatrix}.
\end{equation}
Using these results, equation (\ref{eq:EOM_S_matter}) gives
\begin{equation}
    \mathcal{D}\left[e^{-\Phi}\sum_{k=1}^{\infty}\alpha'^{k-1}4k c_k (-1)^{k-1}2^{2k-1}\begin{pmatrix}
    -H_i^{2k-1} & 0\\
     0 & H_i^{2k-1} \end{pmatrix}\right] = -\kappa^2\sqrt{g}\begin{pmatrix}
    -p_i & 0\\
     0 & p_i \end{pmatrix},
\end{equation}
which implies that
\begin{equation}
    \mathcal{D}\left(e^{-\Phi}\sum_{k=1}^{\infty}(-\alpha')^{k-1}4k c_k 2^{2k-1}H_i^{2k-1}\right) = -\kappa^2 \bar{p}_i.
\end{equation}
In terms of $F(H_i)$ we have
\begin{equation}
    \mathcal{D}\left(e^{-\Phi}F'(H_i)\right) = -2d \kappa^2 \bar{p}_i,
\end{equation}
which can be also written as
\begin{equation}
    (\mathcal{D}H_i) F''(H_i) - (\mathcal{D}\Phi) F'(H_i) = -2d \kappa^2 e^{\Phi} \bar{p}_i.
\end{equation}

Summarizing, the equations of motion coming from single-trace operators in the action for the anisotropic ansatz (\ref{nonisoansatz}) are
\begin{subequations}\label{singletraceeqs}
    \begin{align}\label{singletraceeqs1}
    2 \mathcal{D}^2\Phi - (\mathcal{D}\Phi)^2 + \frac{1}{d}\sum_{i=1}^d F(H_i) &= \kappa^2 e^{\Phi}\bar{\sigma}, \\
    (\mathcal{D}\Phi)^2 + \frac{1}{d}\sum_{i=1}^d \left(H_i F'(H_i) -F(H_i)\right) &= 2\kappa^2 e^{\Phi}\bar{\rho}, \label{singletraceeqs2}\\
    \mathcal{D}\left(e^{-\Phi}F'(H_i)\right) &= -2d \kappa^2 \bar{p}_i. \label{singletraceeqs3}
\end{align}
\end{subequations}

The continuity equation is not independent of equations $(\ref{eq:EOM_matter})$, as expected from Bianchi identities and explicitly shown in \cite{Bernardo:2019bkz}. It is written as 
\begin{equation}
    \dot{\bar{\rho}} + \frac{1}{4}\text{tr}(\mathcal{S}\dot{\mathcal{S}}\eta \bar{\mathcal{T}}) - \frac{1}{2}\bar{\sigma}\dot{\Phi} = 0,
\end{equation}
and given the anisotropic ansatz it reduces to
\begin{equation}
    \dot{\bar{\rho}} + \sum_{i=1}^d H_i \bar{p}_i - \frac{1}{2}\bar{\sigma}\dot{\Phi} = 0.
\end{equation}

In Appendix \ref{check}, we show that upon neglecting $\alpha'$-corrections equations (\ref{singletraceeqs}) reduce to the dilaton-gravity equations coupled to matter. This was to be expected since the lowest order action has no contribution from multi-trace operators and equations (\ref{singletraceeqs}) contain only contributions from the single-trace operators. Despite of this fact, we show how and why they are still relevant for discussing anisotropic solutions in next subsection.
 
\subsection{Non-isotropic solutions with static internal directions}\label{sec:non_isotropic_solutions}

Recalling the discussion after equation (\ref{singleandmultitraceopunderansatz}), if we impose $a_i(t) = a(t)$ for $(d-n)$ directions and let the other $n$ directions to have constant scale factor, $H_i = 0$ for $i=d-n+1, \cdots, d$, then the multi-trace operators can be neglected (in the sense that they will contribute in the same manner as the single-trace ones), and then the equations of motion (\ref{singletraceeqs}) are the full set of equations including all $\alpha'$-corrections. In this section, we will obtain a consistency condition that the matter sector should satisfy in order to have $H_i = 0$ for $n$ directions. 

Let us rewrite (\ref{singletraceeqs}) for $(d-n)$ isotropic directions with the same scale factor $a(t)$ and corresponding Hubble rate $H(t)$ (setting $n(t)=1$):
\begin{subequations}
\begin{align}
    2\Ddot{\Phi} - \dot{\Phi}^2 +\frac{(d-n)}{d}F(H) + \frac{1}{d}\sum_{i=d-n+1}^{d}F(H_i) &= \kappa^2 e^{\Phi} \bar{\sigma},\\
    \dot{\Phi}^2 + \frac{(d-n)}{d}(H F'(H)-F(H)) + \frac{1}{d}\sum_{i = d-n+1}^{d}(H_i F'(H_i)- F(H_i)) &= 2\kappa^2 e^{\Phi}\bar{\rho},\\
    \partial_t\left(e^{-\Phi} F'(H)\right) &= -2 d \kappa^2 \bar{p},\\
    \partial_t\left(e^{-\Phi} F'(H_i)\right) &= -2 d \kappa^2 \bar{p}_i,
\end{align}
\end{subequations}
where the index $i$ in the last equation runs over the $n$ directions with different scale factors $a_i$ ($i = d-n+1, \cdots, d$). From the last equation we see that we need $p_i = 0$ in order to have static internal directions. So any solution with zero pressure in $n$ different directions and with same scale factor $a(t)$ for $(d-n)$ directions satisfying
\begin{subequations}\label{eq:anisotopric_alpha_cosm_F}
    \begin{align}
        \label{d-ndimeq1}
    2\Ddot{\Phi} - \dot{\Phi}^2 +\frac{(d-n)}{d}F(H) &= \kappa^2 e^{\Phi} \bar{\sigma}\\
    \dot{\Phi}^2 + \frac{(d-n)}{d}(H F'(H)-F(H)) &= 2\kappa^2 e^{\Phi}\bar{\rho}\label{d-ndimeq2}\\
    \partial_t\left(e^{-\Phi} F'(H)\right) &= -2 d \kappa^2 \bar{p}
    \end{align}
\end{subequations}
\emph{is a solution for the entire set of corrected equations, with $n$ stabilized directions.} For this specific class of solutions, we can use the results from $\alpha'$-Cosmology even though they do not include multi-trace contributions. 

Indeed, rather than trying to find new solutions for the (new) set of equations \eqref{eq:anisotopric_alpha_cosm_F}, we can use a trick to map the new equations to the ones in \cite{Bernardo:2019bkz,Bernardo:2020zlc}. The trick is the following: if we define a function $J(x)$ as
\begin{equation}
    J(x) \equiv \frac{d-n}{d}F(x),
\end{equation}
then we can write \eqref{eq:anisotopric_alpha_cosm_F} as
\begin{subequations}\label{eq:anisotopric_alpha_cosm}
    \begin{align}
    2\Ddot{\Phi} - \dot{\Phi}^2 +J(H) &= \kappa^2 e^{\Phi} \bar{\sigma} \\
    \dot{\Phi}^2 +H J'(H)-J(H) &= 2\kappa^2 e^{\Phi}\bar{\rho} \\
    \partial_t\left(e^{-\Phi} J'(H)\right) &= -2 (d-n) \kappa^2 \bar{p}
\end{align}
\end{subequations}
and the resulting set of equations is the same as the the ones studied in the previous papers once we do the replacements\footnote{We also need to change the $c_k$ coefficients of $F(H)$ to new ones $c'_k$ due to the inclusion of the multitrace contributions. In the rest of the text, we simply drop the prime and continue to denote the set of unknown coefficients by $\{c_k\}$.} $F(H) \rightarrow J(H)$ and $d \rightarrow (d-n)$. \emph{This means that the analyses and solutions in} \cite{Bernardo:2019bkz,Bernardo:2020zlc} \emph{can be used, but now as $(d-n)$-dimensional results!} In particular, there are solutions with $H = \text{const.}$ with EoS $w = p/\rho = -1/(d-n)$ and a perturbative solution with constant dilaton that asymptotes to a $(d-n)$-dimensional spatial volume dominated by radiation.

\section{Emergent Cosmological Scenario with four large dimensions}\label{the_model} 
Having at our hands all the necessary equations and solutions to build our cosmological model, now we focus on its details. We will first highlight how the dynamics happens in the String frame and later describe  the corresponding picture in the Einstein frame. 

\subsection{Dynamics in the String Frame}\label{String_Frame}

\subsubsection{Stage 1 - winding EoS: $w^{(1)} = - 1/d$} \label{SF_stage_1}

We start off with a homogeneous and isotropic spacetime with topology $\mathbb{R} \times \mathbb{T}^d$ where all the dimensions\footnote{Note that although we consider $D = 1 + d = 1 + (d-n) + n = 10$, where $(d-n)$ corresponds to three large dimensions and $n$ to six small dimensions, we keep the notation general.} are smaller than the string length, $l_s = \ap^{1/2}$. Thus, 
\begin{equation}
    R(t) \equiv r_c a(t) < l_s,
\end{equation}
where $R(t)$ corresponds to the size of the compact dimensions and $r_c$ is the comoving radius of them. Looking to the mass spectrum of closed strings in a compact space, we know that the dominant modes for when $R(t) < l_s$ are the winding modes\footnote{Postulating T-duality to be true, we know that that has to be the case since for a large radius we only have momentum modes being excited, which transform into winding modes when T-dual rotated.}\cite{Battefeld:2005av} with corresponding EoS given by $w^{(1)}=-1/d$. We label each different stage from now on with a superscript $(i)$, where $i=1,2 \text{ and } 3$.

Given that the directions' sizes are within the string scale, the non-perturbative equations that take into account the infinite tower of $\ap$-corrections are suitable to find solutions for such regime. In fact, it has been shown that one solution for this matter content results in a dS universe, thus $H(t) = H^{(1)} = \text{const.}$, with dilaton evolving as \eqref{eq:dot_dilaon_for_dS},
\begin{equation}
    \phi^{(1)}(t) = \phi_0 + \frac{d-1}{2} H^{(1)} t, \quad t \geq 0 \label{eq:dilaton_phase_1},
\end{equation}
which grows linearly in time. It is easy to see from \eqref{HEnoniso} that this solution corresponds to having all the dimensions being static in the EF. 

Hence, all the dimensions are exponentially growing until they reach the string scale at the time $t_s$, representing the end of the first phase. That happens when, 
\begin{equation}
    t_s = \frac{1}{H^{(1)}} \ln \left(\frac{l_s}{a_0 r_c} \right), \label{eq:end_1st_phase}
\end{equation}
where $a_0$ is a constant corresponding to the initial value of the scale factor which can be absorbed into $r_c$, which now becomes the initial size of the dimensions in the SF. 

Since the dilaton is growing in time, we need to track its evolution to avoid entering too early into the strong coupling regime where quantum corrections can no longer be neglected. At the end of stage 1, the dilaton is
\begin{equation}
    \phi^{(1)} (t_s) = \phi_0 + \frac{d-1}{2} \ln \left( \frac{l_s}{r_c} \right), \label{eq:dil_end_phase_1}
\end{equation}
and so, in order to have $e^{\phi^{(1)}(t_s)}< 1$, we need the initial value of the string coupling to satisfy $e^{\phi_0}<(r_c/l_s)^{(d-1)/2}$. Note that, as we would like to start in the weak coupling regime, this condition is compatible with the assumption that $r_c< l_s$. In other words, starting with all directions compactified with physical radius smaller than the string length and requiring the string coupling to be smaller than one at $t_s$ \emph{implies} that the string coupling is small at the initial time.

Once the size of the dimensions have reached the string length, we expect another solution to be relevant to our dynamics. In particular, we know that the equation of state cannot be given purely by the winding EoS until the end of this stage and beyond, since as the size of the dimensions grows other modes are excited: oscillatory and momentum modes. In fact, we expect that at the end of this phase the EoS is $w=0$ \cite{Brandenberger:1988aj,Battefeld:2005av}. However, it is very hard to solve the equations for an evolving EoS, and so our scenario will be built by gluing together solutions of three different stages while describing what is expected to be happening in between them. We start with the first transition now.

\subsubsection{Transition between stage 1 and stage 2}\label{SF_transition_1_2}

As the sizes of the dimensions are growing and the EoS is evolving towards zero due to the excitation of oscillatory and momentum modes, we expect the expansion to slow down and we exit the dS phase. That means that during the transition between stage 1 and stage 2 we have that\footnote{See Appendix \ref{appendixA}.} $\dot{H}^{(1)\rightarrow(2)}<0$ while we keep $H^{(1)\rightarrow(2)}>0$. At the same time, we know that the static phase in the EF has to end as the equation of state is evolving, which results in $\dot{H}_E^{(1)\rightarrow(2)}(t_E)>0$ \cite{Brandenberger:1988aj}, which from \eqref{HEnoniso} implies the condition
\begin{equation}
\frac{1-d}{2}\dot{H} + \frac{\dot{\phi}^2}{2(d-1)} < \dot{\phi}H-\ddot{\phi},
\end{equation}
where the LHS is positive, thus
\begin{equation}
    \frac{d \ln \dot{\phi}}{d \ln a} < 1, \label{eq:dil_ph1_to_ph2}
\end{equation}
where we have suppressed the label ``$_{(1)\rightarrow(2)}$''. Thus, a ``slow-rolling" dilaton during the transition between the two phases is a necessary condition for having $H_E^{(2)}>0$. We do not know how fast this transition is but it could be modelled phenomenologically and numerically through \cite{Brandenberger:2018xwl}
\begin{equation}
    w(a) = \frac{2}{\pi d}\arctan ( \gamma \ln{a} ),
\end{equation}
where $\gamma$ indicates how fast the transition happens.

\subsubsection{Stage 2 - matter EoS, $w^{(2)}=0$}\label{SF_stage_2}

As the dimensions reach the string size, the dynamics starts to become more involved. Although the EoS is $w=0$, meaning the pressure is vanishing for all the dimensions, we do not expect that isotropy remains for long. The reason for that is that winding modes can annihilate completely in three spatial dimensions \cite{Brandenberger:1988aj} (from now on called external directions, labelled $d-n=3$ and with scale factor $a(t)$) while they continue to exist for the remaining six other dimensions (from now on called internal directions, labeled $n=6$ and with scale factor $b(t)$).

The internal directions start to oscillate around the self-dual radius due to the interplay between winding and momentum modes, which in average results in $w^{(2)}_b=0$, until they eventually stabilize, $H^{(2)}_b=0$ \cite{Battefeld:2005av}. It is exactly this dynamics that justifies the simplification considered above in Section \ref{sec:non_isotropic_solutions}, where the multi-trace terms were considered zero for an anisotropic universe since all the internal directions were considered static, allowing a decoupling between the dynamics of the internal and external directions. Thus, this is the end of the dynamics of the internal directions in the SF. 

Turning now to the external directions, the phase with EoS parameter close to zero is very short, since $\dot{w}^{(2)}_a > 0$ even though $w^{(2)}_a \approx 0$. Thus, we cannot describe this phase by considering $H^{(2)}_a=0$ as we did for the internal directions. Although the Hubble parameter for these dimensions decreases as the oscillatory and momentum modes are excited, the winding modes will start annihilating and the EoS rapidly converges to a radiation EoS as the expansion continues. 
To describe this short stage, we consider as a first order approximation $H_a^{(2)}$ to be constant and smaller than $H^{(1)}_a=H^{(1)}$ while $w^{(2)} \approx 0$. The dilaton evolves as \eqref{eq:dot_dilaon_for_dS} with $w_a = 0$. Thus, 
\begin{equation}
    \phi^{(2)}(t) = \phi^{(1)}(t_s) + \frac{d-n}{2} H^{(2)} (t - t_s) \label{eq:dilaton_phase_2}, \quad t\geq t_s
\end{equation}
where we have glued\footnote{Note that we do not glue derivatives since we do not consider a continuous evolution of the EoS.} the two solutions after imposing $\phi^{(2)}(t_s)= \phi^{(1)}(t_s)$. Note that \eqref{eq:dilaton_phase_2} shows the dilaton's evolution depending only on the external directions (see discussion on Section \ref{sec:non_isotropic_solutions}), which is now evolving slower as expected from \eqref{eq:dil_ph1_to_ph2}. We will consider this solution to be valid until all the winding modes annihilate and the EoS of the external directions becomes radiation, at time $t_r$. So, 
\begin{equation}
     \phi^{(2)}(t_r) = \phi^{(1)}(t_s) + \frac{d-n}{2} H^{(2)} (t_r - t_s),
\end{equation}
at the end of this stage.

\subsubsection{Transition between stage 2 and stage 3}

Stage 2 is rather short  since  the external directions continue to be exponentially expanding, implying that the winding modes continue to decay away while the momentum modes are being excited and the EoS rapidly approaches the one of radiation, $w=1/(d-n)$. Thus, this transition is characterized by having a positive EoS with $\dot{w}>0$. 

Unfortunately, due to the unknown coefficients $c_k$, we cannot solve the equations for $H$ during the transition. We will assume that the rate of expansion does not change considerably, remaining around the string scale so that non-perturbative solutions will have to be invoked during the next stage. We will come back to this point in Section \ref{Conclusion}.

\subsubsection{Stage 3 - radiation EOS, $w^{(3)}_a = 1/(d-n)$}\label{SF_stage_3}

\paragraph{Rolling dilaton.} For this stage we consider that the EoS is already given by that of radiation for the external directions. Note that for the low-energy theory that automatically implies that the dilaton is constant (see e.g., \cite{Tseytlin:1991xk}), while for $\ap$-Cosmology that is not necessarily the case, especially around the string scale. Initially the dilaton keeps evolving following \eqref{eq:dot_dilaon_for_dS} with $w_a = 1/d$
\begin{equation}
    \phi^{(3)}(t) = \phi^{(2)}(t_r) + \frac{d-n+1}{2} H^{(3)} (t - t_r) \label{eq:dilaton_phase_3}, \quad t\geq t_r
\end{equation}
where we once again glued solutions imposing $\phi^{(3)}(t_r) = \phi^{(2)}(t_r)$. 
As the expansion continues and the dilaton keeps evolving, at some point the dilaton will approach zero, leading the dynamics into the strong regime, since $g_s =e^\phi \sim 1$. We invoke its stabilization during this regime. We do not have a precise dynamics to account for that, we can only say that in the strong regime quantum corrections to the dilaton's dynamics can become important, resulting in an effective potential that stabilizes it to a constant value \cite{Gasperini:2007ar, Damour:1994zq,Danos:2008pv}. 

Note that the time $t_*$ when the strong regime is reached (which we assume to be when the dilaton reaches the value zero) in our scheme depends on the initial value $\phi_0$ of the dilaton, the string scale, the initial size of the dimensions, and the Hubble parameters at the different stages. The approximate relation is
\begin{equation}
      t_* H^{(3)} (d-n+1) + t_r [(d-n) H^{(2)} - (d-n+1) H^{(3)}] - t_s [H^{(2)} (d-n) - H^{(1)} (d-1)]  \approx 2|\phi_0| \, .
\end{equation}
This is a constraint on the parameters introduced so far so to ensure that the strong regime is approached only at stage 3. In fact, we can approximate this expression considering that $t_r \approx t_s$, since stage 2 is very short. Then, we obtain 
\begin{equation}
    t_* H^{(3)} (d-n+1) - t_s [(d-n+1) H^{(3)}  - (d-1) H^{(1)}] \approx 2|\phi_0|.
\end{equation}
After this transition, $t>t_*$, both String and Einstein frames become completely equivalent. Moreover, now that the dilaton is halted the internal directions are also completely stabilized in the EF.  

\paragraph{Constant dilaton.} Once the dilaton is constant and we progressively leave the string scale, perturbative solutions can be used. Solutions of this kind have been found in \cite{Bernardo:2019bkz} and they take the form 
\begin{align}
    H_a^{(3)}(t) &= \frac{2}{(d-n+1)t} + \alpha' \frac{H_1}{t^3} + \alpha'^2 \frac{H_2}{t^5} + \dots, \label{eq:pert_sol_const_dil}\\
    w^{(3)}_a(t) &= \frac{1}{d-n}-32 (d-n)  c_2 w_2 \alpha' H^2 +128 (d-n) c_3 w_3 \alpha'^2 H^4 -512 (d-n) c_4 w_4 \alpha'^3 H^6 + \dots, \ \label{eq:pert_sol_const_dil_EOS}
\end{align}
where the EoS for the external directions also evolves towards radiation, $w_a = 1/(d-n)$. Thus, in principle one could consider a smooth gluing between the last two stages. As the expansion continues, all the corrections decay away and we recover the known radiation solution of the lowest order theory, for which the Hubble parameter of the external directions evolves purely as radiation while the dilaton is completely fixed.

\subsection{Dynamics in the Einstein frame} \label{subsec:Einstein_frame}

Apart from the perturbative solution with radiation EoS and the static phase in the SF with matter EoS, all other solutions considered are dS solutions in the SF with different equations of state (and, consequently, different dilaton evolution). Moreover, the  late time perturbative solution has a constant dilaton and its EF dynamics is identical to the SF one. Thus, we will be interested in the $H_E(t_E)$ for the dS solutions in the following.

As we are considering a non-isotropic ansatz, we need to rederive the formula for $H_E(t_E)$ in terms of the String frame variables. The Einstein frame metric is given by
\begin{equation}
    G^{E}_{\mu\nu} \equiv e^{-\frac{4\phi}{d-1}}G_{\mu\nu}
\end{equation}
and so the relation between the time variables in both frames is the same as in the isotropic case. Focusing now on the spatial components, we have that 
\begin{equation}
    a^2_{i, E}(t) = e^{-\frac{4\phi}{d-1}}a^2_i, 
\end{equation}
and starting from this it is straightforward to show that
\begin{equation}\label{HEnoniso}
    H_{i,E}(t_E) = - \frac{e^{\frac{2\phi}{d-1}}}{d-1}\left(\dot{\Phi}+ \sum_{j=1}^d H_j - (d-1)H_i \right) = -\left(\prod_{l=1}^d a_l^{\frac{1}{d-1}}\right)\frac{e^{\frac{\Phi}{d-1}}}{d-1}\left(\dot{\Phi}+ \sum_{j=1}^d H_j - (d-1)H_i \right).
\end{equation}

Now, let us write the dS solutions in the SF in terms of the Einstein variables. For such solutions (including a dilatonic charge), it was shown in \cite{Bernardo:2020zlc} that
\begin{equation}
    H_i = H_0, \quad \dot{\Phi} = -\beta H_0.
\end{equation}
The relation between $\beta$ and $(w, \lambda)$ depends on the spacetime dimensionality. Considering $n$ stabilized directions with $H_b = 0$, following the discussion in Section \ref{sec:non_isotropic_solutions}, we have
\begin{equation}
    \beta = -\frac{(d-n)w}{1 + \lambda/2},
\end{equation}
and the dilaton evolution is
\begin{equation}\label{dilatonfordS}
    \phi(t) = \phi_0 + \frac{(d-n -\beta)}{2}H_0 (t - t_0),
\end{equation}
since only the evolving directions contribute to $\dot{\Phi}$. Note that in order to write the non-isotropic solution with $n$ stabilized directions, we are invoking the arguments of Section \ref{sec:non_isotropic_solutions}, which allows us to have a solution in the String frame that is locally $\text{dS}_{d-n}\times \mathbb{T}^n$.

The time variable in the EF can be obtained as
\begin{equation}
    dt_E = e^{-\frac{2\phi}{d-1}}dt \implies t_E - t_{E,0} = e^{-\frac{2\phi_0}{d-1}}\frac{d-1}{(d-n-\beta)H_0}\left(1- e^{-\frac{d-n-\beta}{d-1}H_0 (t - t_0)}\right),
\end{equation}
where $t_{E,0}$ is the value of $t_E(t)$ at $t=t_0 
$. We can invert this result to write $t(t_E)$:
\begin{equation}
    t - t_0 = -\frac{d-1}{(d-n-\beta)H_0}\ln\left[1 - e^{\frac{2\phi_0}{d-1}}\frac{(d-n-\beta)H_0}{d-1}(t_E - t_{E,0})\right].
\end{equation}
The Einstein frame Hubble parameter $H_E(t_E)$ for the $(d-n)$ external directions is obtained from (\ref{HEnoniso}) evaluated for the dS solution,
\begin{equation}
    H_E(t_E) = e^{\frac{2\phi_0}{d-1}}H_0\frac{(\beta +n -1)}{d-1}\left[1 - e^{\frac{2\phi_0}{d-1}}\frac{(d-n-\beta)}{d-1}H_0 (t_E - t_{E,0})\right]^{-1}.
\end{equation}
Note that as the bracket in the above expression came from an exponential, it cannot be negative. We can check that while $t \in [0, \infty)$, the relation $t_E(t)$ is such that $t_E \in [t_{E,0}, t_{E,\text{max}} )$ where
\begin{equation}
    t_{E,\text{max}} =  t_{E,0} + e^{-\frac{2\phi_0}{d-1}}\frac{d-1}{(d-n-\beta)H_0},sd
\end{equation}
such that the bracket in $H_{E}(t_E)$ is never negative. Since for a gas of strings we can set $\lambda = 0$, we have $\beta = -(d-n)w$, such that
\begin{equation}\label{HEfordS}
    H_E(t_E) =  e^{\frac{2\phi_0}{d-1}}\frac{(n-1-(d-n)w)}{d-1}H_0 \left[1 - e^{\frac{2\phi_0}{d-1}}\frac{(d-n)(w+1)}{d-1}H_0 (t_E - t_{E,0})\right]^{-1}.
\end{equation}

The evolution of the dilaton in the EF is simply
\begin{equation}
    \phi(t_E) = \phi_0 - \frac{d-1}{2}\ln\left[1 - e^{\frac{2\phi_0}{d-1}}\frac{(d-n-\beta)H_0}{d-1}(t_E - t_{E,0})\right],
\end{equation}
from which we can see that it is never singular. Moreover, the canonically normalized scalar field associated to the dilaton is given by (see for instance \cite{Polchinski:1998rq})
\begin{equation}
    \varphi(t_E) = \frac{2}{\kappa}\frac{1}{\sqrt{d-1}}\phi(t_E),
\end{equation}
and clearly has non-singular evolution for dS solutions in the SF.

\subsubsection{Stage 1 - winding EoS: $w^{(1)} = - 1/d$} \label{EF_stage_1}

During this phase all dimensions are evolving with the same scale factor and winding EoS. Thus, we can use equation (\ref{HEfordS}) with $n=0$ and $w =-1/d$, yielding
\begin{equation}
    H_E^{(1)}(t_E) = 0,
\end{equation}
as first realized in \cite{Bernardo:2020zlc}. Therefore, we start off with a 10-dimensional static phase in Einstein frame.

\subsubsection{Stage 2 - matter EoS, $w^{(2)}=0$}\label{EF_stage_2}

During this stage, we have a matter EoS in all directions, i.e., $p_i = 0$ for all $i$, but the dynamics is different for the internal and external directions. For the former, $H_b^{(2)} = 0$ and for the latter, $H_a= H^{(2)}=\text{constant}$. The dilaton time dependence is still fixed by the evolution of the external directions and is given by equation (\ref{dilatonfordS}). For the internal directions, equation (\ref{HEnoniso}) gives
\begin{equation}
    H_{b, E}(t_E) = -\frac{e^{\frac{2\phi}{d-1}}}{d-1}[\dot{\Phi} + (d-n)H_a],
\end{equation}
and evaluating this for the dS solution, we get
\begin{equation} \label{eq:HE_internal}
    H_{b, E}^{(2)}(t_E) = -\frac{e^{\frac{2\phi^{(2)}}{d-1}}}{d-1}H^{(2)}(d-n - \beta).
\end{equation}
Thus, for $w=0$ ($\beta = 0$), the internal directions are contracting in the EF. They will keep contracting until the dilaton is stabilized at the end of stage 3. 

On the other hand, for the external directions, equation (\ref{HEnoniso}) implies that
\begin{equation}
    H_{a, E}(t_E) = -\frac{e^{\frac{2\phi}{d-1}}}{d-1}(\dot{\Phi} - (n-1)H_a),
\end{equation}
which evaluated for the dS solution gives
\begin{equation}\label{HaEnoniso}
     H_{a, E}^{(2)}(t_E) = \frac{e^{\frac{2\phi^{(2)}(t_s)}{d-1}}}{d-1}H^{(2)}(n-1+ \beta)\left[1 - e^{\frac{2\phi^{(2)}(t_s)}{d-1}}\frac{(d-n-\beta)}{d-1}H^{(2)} (t_E - t_E(t_s))\right]^{-1},
\end{equation}
that is, for $w=0$, the external directions are expanding with rate
\begin{equation}
    H_{a, E}^{(2)}(t_E) = \frac{e^{\frac{2\phi^{(2)}(t_s)}{d-1}}}{d-1}H^{(2)}(n-1)\left[1 - e^{\frac{2\phi^{(2)}(t_s)}{d-1}}\frac{(d-n)}{d-1}H^{(2)} (t_E - t_E(t_s))\right]^{-1}. \label{eq:stage_2_Einstein_sol}
\end{equation}
Note that this corresponds to a  phase of super-exponential accelerated expansion which, if stage 2 were to last a long time, would lead to a finite time singularity. We come back to this point below.

\subsubsection{Stage 3 - radiation EOS, $w^{(3)}_a = 1/(d-n)$}\label{EF_stage_3}

During this stage, before the string coupling becomes $\mathcal{O}(1)$, the SF dynamics of the internal directions is trivial, since they are stabilized with $w_b = 0$, while the external directions are expanding with constant $H_a = H^{(3)}$ and radiation EoS $w_a=1/(d-n)$. Thus, in the EF the internal directions are still contracting as in stage 2, as it can be seen from equation (\ref{eq:HE_internal}) for $w = 1/(d-n)$ (i.e., $\beta =-1$), and the external directions expand as
\begin{equation}
    H_{a, E}^{(3)}(t_E) = \frac{e^{\frac{2\phi^{(3)} (t_r)}{d-1}}}{d-1}H^{(3)}(n-2)\left[1 - e^{\frac{2\phi^{(3)}(t_r)}{d-1}}\frac{(d-n+1)}{d-1}H^{(3)} (t_E - t_E(t_r))\right]^{-1},
\end{equation}
as can be checked from (\ref{HaEnoniso}). This phase ends when the dilaton value is such that quantum effects cannot be neglected anymore and non-perturbative loop effects might dominate the dynamics. Then, such effects could stabilize $\phi$ such that there is no difference between the frames anymore.

Similarly to \eqref{eq:stage_2_Einstein_sol}, this solution also corresponds to a phase of accelerated expansion with a finite time singularity. Fortunately, both stages are short in our model, since the former describes the fleeting phase where the EoS is approximately zero while the latter models the end of the non-perturbative dynamics in $\ap$ before the dilaton is fixed. Nonetheless, these two phases might play an important role to explain away the flatness and horizon problems present in the Standard Big Bang Cosmology. This is further explored in \cite{Bernardo:2020bpa}.

 Once the dilaton is stabilized, the string gas with radiation EoS can support the perturbative solution \eqref{eq:pert_sol_const_dil} of \cite{Bernardo:2019bkz}. When this happens, the value of $H$ decreases with time and the tower of $\alpha'$-corrections become more and more irrelevant as the expansion continues to unfold until we finally get a 4-dimensional low-curvature regime dominated by radiation.

\section{Conclusion and Discussions}\label{Conclusion}

In this paper we have built the first very early universe cosmological model based on $\ap$-Cosmology and inspired by the String Gas Cosmology (SGC) scenario. Our model provides for the first time dynamics for the Einstein frame quasi-static phase advocated by SGC. Moreover, with reasonable assumptions, we have shown that the non-perturbative equations of $\alpha'$-Cosmology are compatible with the dynamical mechanism of SGC to generate a $4$-dimensional cosmology starting from ten dimensions, as required by String Theory.

Our dynamical system consists of the equations of $\ap$-Cosmology coupled to a matter sector being given by a gas of strings described by a barotropic perfect fluid. From the thermodynamics of the strings, we can model the evolution of the equation of state both for the internal and external directions. To solve these equations, we break the time evolution into different stages and consider each stage separately.

In our model, all nine spatial dimensions start off with an equal size smaller than the string length, which implies that the dominant modes are winding, with EoS $w=-1/9$. This corresponds to a de Sitter expansion in the String frame and to a static phase in the Einstein frame. As the dimensions expand in the former, the density of states of winding modes decays as other modes are excited, and the EoS grows until it becomes that of a presureless fluid.

As the matter energy drifts from the winding modes to other string excitations and the equation of state parameter approaches $w = 0$, the dynamics stops being isotropic since winding modes can completely disappear only in three spatial dimentions. Thus, there result two sectors, each of which we model as isotropic: one with six internal directions and the other with three external ones. The pressure in the former remains around zero due to the interplay of winding and momemtum modes, while the EoS for the latter keeps evolving towards a radiation EoS, i.e., the density of states is dominated by momentum modes. The internal dynamics freezes out completely in the String frame at this point.

Once the EoS paramter becomes $w=1/3$ for the external directions, the dynamics is divided in two phases: a non-perturbative solution in $\ap$ for when the energy scale is still around the string scale and a perturbative solution for the low-curvature regime. The transition between these two is given by the stabilization of the dilaton. The first phase is described by a short dS solution in the SF, which corresponds to super-exponential acceleration in the EF, while the latter converges to a typical radiation dominated solution for both frames.

Needless to say, our model can be further improved, even within the framework of $\ap$-Cosmology and given that we are at the moment not able to solve the equations for an evolving EoS parameter. In particular, it would be interesting to explore the phenomenological consequences of our result that there is a short phase of accelerated expansion for the external dimensions in the Einstein frame. Could this be enough to account for the observed spatial flatness of the universe (see also \cite{Kamali:2020drm} for a recent attempt to explain the spatial flatness in the context of SGC)? In \cite{Bernardo:2020bpa}
we have shown that the answer is yes, proving that the model can be made compatible with standard cosmology at the background level.

Besides, in order to make contact with the most important cosmological successes of SGC, namely its prediction of almost scale-invariant power spectra for scalar and tensor perturbations with red- and blue-tilts, respectively, we would need to consider cosmological perturbations starting in a ten-dimensional isotropic background with a rolling dilaton. The calculations considered in the context of SGC so far have been made for a constant dilaton and in a four-dimensional space. One reason to believe that the results could be robust is the fact that the results of \cite{Nayeri:2005ck,Brandenberger:2006xi} are based mostly on holographic scaling of thermodynamic fluctuations, and this may be robust to the change in the background dynamics.

The reader might have become suspicious about our arguments concerning dilaton stabilization. Indeed, our discussion concerning how this happens remains to be improved in the context of $\ap$-Cosmology. In fact, it may be the case that we do not even need to rely on its stabilization by any other mechanism than the sole evolution of the equation of state. The reason for that is that we might be able to make a strong argument purely based on the equations of motion that imply that $H$ is decaying as the EoS evolves with $\dot{w}>0$ (similarly as derived in Appendix \ref{appendixA} when $w$ is evolving away from the winding mode dominance). Then, the transition between stages 2 and 3 could end with the Hubble parameter already lower than the string scale, such that we could consider directly the perturbative solution with a rolling dilaton for which its evolution asymptotes to a constant \cite{Bernardo:2020xxy}.

Finally, let us comment on the connection of our work to Double Field Theory (DFT) (see e.g. \cite{Aldazabal:2013sca, Hohm:2013bwa} for reviews). The dynamical equations of $\alpha'$-Cosmology as studied in \cite{Hohm:2019jgu, Bernardo:2019bkz, Bernardo:2020zlc} do not necessarily assume a compact background. Thus, the O$(d,d)$ group explored in those works is present even in the non-compact case. In our model, we have assumed a compact background, so the O$(d,d)$ discussed in the present work is part of the T-duality group. In fact, we can recover it from the generalized coordinate transformations of DFT, see for instance \cite{Hohm:2013bwa}. Thus, a possible avenue of exploration is to embed our model into DFT, or at least to describe its first stage in a T-dual frame, where instead of considering the directions' size to be smaller than the string scale and expanding, the dimensions are large and contracting, cf. \cite{Brandenberger:2018bdc,Bernardo:2019pnq}.

While the present work was in review, an interesting paper \cite{Nunez:2020hxx} appeared presenting new vacuum solutions including a non-trivial NS-NS 2-form field. In the model developed here, the energy density of the string gas source cannot be neglected in any phase. Hence,
the new solutions of \cite{Nunez:2020hxx} cannot be immediately used to improve our model. However, the equations developed in \cite{Nunez:2020hxx} are more general than the ones used to get the solutions of section \ref{technical}, for they 
include the coupling with the $B_{\mu \nu}$ field. It would be
of great interest to add matter to the setup of \cite{Nunez:2020hxx} and to study
whether this would help us improve our model.

\section*{Acknowledgments}

The research at McGill is supported, in part, by funds from NSERC and from the Canada Research Chair program. RB thanks the  Pauli Center and the Institutes for Theoretical Physics and of Particle Astrophysics of the ETH for hospitality.

\appendix

\section{Transition between Stage 1 and Stage 2: $\dot{H}<0$}\label{appendixA}

We need to show that as the EoS increases over time, $\dot{w}>0$ and while it remains negative, $w<0$, the Hubble parameter decreases when starting off in a dS phase. To study what conditions are necessary for this to happen, we consider linear perturbations of the equations \eqref{eq:anisotopric_alpha_cosm_F} by introducing
\begin{subequations}
    \begin{align}
    H(t) &= H_0 (t) + H_1(t)\\
    \dot{\Phi}(t) &= \dot{\Phi}_0 (t) + \dot{\Phi}_1 (t)\\
    \bar{\rho}(t) &= \bar{\rho}_0 (t) + \bar{\rho}_1(t)\\
    \bar{p}(t) &= \bar{p}_0 (t) + \bar{p}_1(t),
\end{align}
\end{subequations}
where the subscript ``${_1}$'' denotes the perturbations. Plugging this ansatz into \eqref{eq:anisotopric_alpha_cosm_F} with $n=0$ and using the background equations, the resulting first order equation for the perturbations are
\begin{subequations}\label{sub:perturbed_eqs}
\begin{align}
        2 \dot{\Phi}_0 \dot{\Phi}_1 +  H_0 F''_{0}H_1 &= 2\kappa^2 e^{\Phi_0} (\bar{\rho}_1 + \bar{\rho}_0 \Phi_1)\label{sub:perturbed_eqs_a} \\
        F''_{0} \dot{H}_1 + \left[\dot{H}_0 F^{(3)}(H_0) - \dot{\Phi}_0 F''_{0}\right] H_1 - F'_{0} \dot{\Phi}_1 &= - 2 \kappa^2 d e^{\Phi_0} (\bar{p}_1 + \bar{p}_0 \Phi_1) \label{sub:perturbed_eqs_b} \\
        \ddot{\Phi}_1 - \dot{\Phi}_0 \dot{\Phi}_1 +  \frac{1}{2} F'_{0} H_1 &= 0, \label{sub:perturbed_eqs_c} 
  \end{align}
\end{subequations}
and we can also consider the perturbed continuity equation,
\begin{equation}
    \dot{\bar{\rho}}_1 + d H_0 \bar{p}_1 + d \bar{p}_0  H_1 = 0.
\end{equation}
Moreover, for a barotropic EoS, we have
\begin{equation}
    \bar{p}_1 = w_0 \bar{\rho}_1 + \bar{\rho}_0 w_1.
\end{equation}

Now we consider our particular backround $\{ H_0(t),\dot{\Phi}_0(t),w_0(t) \} = \{ H_0, - H_0, -1/d\}$, which also implies $F_0' = 2H_0$ and $F_0 = H_0^2$ as shown in \cite{Bernardo:2020zlc}. Thus, combining \eqref{sub:perturbed_eqs_a} and \eqref{sub:perturbed_eqs_b} results in 
\begin{equation}
    F_0'' \dot{H}_1 = -2\kappa^2 d \bar{\rho}_0 e^{\Phi_0} w_1.
\end{equation}
Since $w_1>0$ as the EoS evolves from $-1/d$ to $0$, we know that as long as $F_0''>0$ we will have a decreasing Hubble parameter, $\dot{H}_1<0.$

\section{A small check: dilaton-gravity equations}\label{check} 

In absence of corrections, we do not need to worry about single or multitrace operators: the equations of motion follow only from the first terms in the action. Then, if we neglect the corrections and take only the leading term in $F(H_i) = -d H_i^2 +\dots$, we can recover the matter sourced dilaton-gravity equations from $\alpha'$-Cosmology. 

In order to show this, let us consider $(d-n)$ directions with same scale factor $a(t)$ and $n$ directions with scale factor $b(t)$. In comparing with SGC, we have in mind the example $d =9$ and $n = 6$. Let us also set $n^2(t) = 1$. Then, we have
\begin{subequations}
    \begin{align}
2 \Ddot{\Phi} - \dot{\Phi}^2 + \frac{1}{d}\left[(d-n)F(H_a) + n F(H_b)\right] &= \kappa^2e^{\Phi}\bar{\sigma},\\
\dot{\Phi}^2 + \frac{d-n}{d}\left(H_a F'(H_a) - F(H_a)\right) + \frac{n}{d}\left(H_b F'(H_b) - F(H_b)\right) &= 2\kappa^2e^{\Phi}\bar{\rho},\\
\partial_t\left(e^{-\Phi}F'(H_a)\right) &= -2 d\kappa^2 \bar{p}_a,\\
\partial_t\left(e^{-\Phi}F'(H_b)\right) &= -2 d\kappa^2 \bar{p}_b,
\end{align}
\end{subequations}
where we denote $H_a= d\ln a/dt$, $H_b = d\ln b / dt$ while $p_a$ and $p_b$ the respective pressures. Putting $\sigma = 0$, neglecting $\alpha'$-corrections by setting $F(x) = -dx^2$ and defining $\lambda = \ln a$ and $\nu = \ln b$ we have
\begin{subequations}
\begin{align}
2\Ddot{\Phi}-\dot{\Phi}^2- (d-n)\dot{\lambda}^2- n \dot{\nu}^2 &= 0,\label{SGCeqsigma}\\
\dot{\Phi}^2- (d-n)\dot{\lambda}^2- n \dot{\nu}^2 &= 2\kappa^2e^{\Phi}\bar{\rho},\label{SGCeqrho}\\
-\dot{\Phi}\dot{\lambda} + \Ddot{\lambda} &= \kappa^2e^{\Phi}\bar{p}_a,\\
-\dot{\Phi}\dot{\nu} + \Ddot{\nu} &= \kappa^2e^{\Phi}\bar{p}_b.
\end{align}
\end{subequations}
We can combine the first two equations to write instead
\begin{equation}
    \Ddot{\Phi} - (d-n) \dot{\lambda}^2 - n \dot{\nu}^2 = \kappa^2e^{\Phi}\bar{\rho}\label{SGCeqsigmaandrho}.
\end{equation}
Given the assumptions of this subsection, the continuity equation reads,
\begin{equation}
    \dot{\bar{\rho}} + (d-n)\dot{\lambda}\bar{p}_a + n \dot{\nu}\bar{p}_b = 0,
\end{equation}
as can be checked by starting with the equations of motion.

These equations match the ones previously used in studies of string cosmology (see, for instance, eqs. (38)-(41) in \cite{Battefeld:2005av} in the absence of flux). There is an important caveat: since we are not necessarily writing the equations in the critical dimension case ($D_{\text{c}}=10$), we should always have added a term in the action proportional to $(D-D_{\text{c}})$:
    \begin{equation}
    \delta S = \frac{1}{2\kappa^2}\int\mathrm{d}^Dx\,\sqrt{-G} e^{-2\phi}\left(-\frac{2}{3\alpha'}(D-D_\mathrm{c})\right)= -\frac{1}{2\kappa^2}\int\mathrm{d}^Dx\,n e^{-\Phi} \Lambda\,.
\end{equation}
However, note that we can simply consider this contribution to the action to be a contribution to the matter action. In this case we would get contributions for the energy density, pressure and dilatonic charge as
\begin{equation}
    \bar{\sigma}_{\Lambda} = -e^{-\Phi}\frac{\Lambda}{\kappa^2}, \qquad \bar{\rho}_{\Lambda} = e^{-\Phi}\frac{\Lambda}{2\kappa^2}, \qquad \bar{p}_{\Lambda} = 0.
\end{equation}
Including these contributions to the matter sector, one can check that the only changes are in equations (\ref{SGCeqsigma}) and (\ref{SGCeqrho}), and they are such that equation (\ref{SGCeqsigmaandrho}) is preserved:
\begin{subequations}
    \begin{align}
    \Ddot{\Phi} - (d-n) \dot{\lambda}^2 - n \dot{\nu}^2 &= \kappa^2e^{\Phi}\bar{\rho},\\
    \dot{\Phi}^2- (d-n)\dot{\lambda}^2- n \dot{\nu}^2 - \Lambda &= 2\kappa^2e^{\Phi}\bar{\rho},\\
    -\dot{\Phi}\dot{\lambda} + \Ddot{\lambda} &= \kappa^2e^{\Phi}\bar{p}_a,\\
    -\dot{\Phi}\dot{\nu} + \Ddot{\nu} &= \kappa^2e^{\Phi}\bar{p}_b.
\end{align}
\end{subequations}
One can also check that the continuity equation is not modified.

\section{Dilaton beta functional and on-shell action}

At lowest order in $\alpha'$, the supergravity action is proportional to the integral of the dilaton equation of motion, which implies that it should vanish on-shell \cite{Callan:1985ia}. Moreover, in \cite{Callan:1986jb} assuming a reasonable form of the fully corrected action, it was shown that the spacetime Lagrangian density should be proportional to the dilaton's conformal anomaly, which implies that the action should vanish to all orders in $\alpha'$. We will check whether this is the case for the non-perturbative SF dS solutions used in this paper assuming a specific action for the matter sector\footnote{We thank the anonymous referee for suggesting such analysis.}.

Upon the cosmological ansatz (\ref{FLRWansatz}), the action (\ref{singletraceaction}) reduces to
\begin{equation}
    S = \frac{1}{2\kappa^2}\int d^dx dt e^{-\Phi}\left[- \dot{\Phi}^2 - F(H)\right] + S_{\text{m}},
\end{equation}
that vanishes for the vacuum solution of \cite{Hohm:2019jgu} due to the constraint on the $F(H)$ function. For the matter coupled case, since a general perfect fluid energy-momentum tensor was employed to find the dS solution (\ref{dSsolution}), we use the Schutz action \cite{Schutz:1970my} to evaluate $S_{\text{m}}$ on-shell\footnote{The Schutz action is valid for any matter action described by a perfect fluid in the level of the equations of motion, which is the case for the matter action consider here.}. Doing so, we get
\begin{equation}
    S_{\text{on-shell}} = \frac{1}{2\kappa^2}\int d^dx dt e^{-\Phi}\left[-(2+\lambda)\beta^2 H_0^2 + 2\kappa^2 e^{\Phi}\bar{p}_{\text{t}}\right],
\end{equation}
where $p_\text{t} = p + \sigma/2$ is the total pressure including the $S_\text{m}$ metric dependence through $\Phi$ \cite{Bernardo:2020xxy}. As a string gas source doesn't depend on $\Phi$, we have $p_\text{t} =p$ for the solutions considered in this paper. The assumption that the Schutz action describes the matter sector doesn't imply that $S_{\text{m}}$ is not invariant under O$(d,d)$ since the pressure can be written as a trace of a term containing the duality covariant energy-momentum tensor $\bar{\mathcal{T}}$.

Now, using the equations of state $p =w   \rho$, $\sigma = \lambda \rho$, the relation between $\rho$ and $H_0$ of \cite{Bernardo:2020zlc} and equation (\ref{relationbetweenEOS}), we find
\begin{equation}
    S_{\text{on-shell}} = \frac{1}{2\kappa^2}\int d^dx dt e^{-\Phi}2(w-1)\beta^2 H_0^2,
\end{equation}
which vanishes for $w =0,\,1$. On the other hand, using the on-shell value of $\Phi$ on the exponential factor we have
\begin{equation}
    S_{\text{on-shell}} = \frac{2\beta H_0 (w-1)}{2\kappa^2}\int d^dx dt \frac{d}{dt}(e^{-\Phi}),
\end{equation}
which shows that the action reduces to a total derivative in time. Thus, we can readily put it to zero since we could have started with an action including a total derivative term to cancel the final on-shell action.

\bibliographystyle{bibstyle} 
\bibliography{references}

\end{document}